# Predicting Non-linear Cellular Automata Quickly by Decomposing Them into Linear Ones

**Cristopher Moore**

*Santa Fe Institute, Santa Fe, New Mexico* moore@santafe.edu

**Timofey Pnin**

*Waindell College, Waindell, New York* timosha@waindell.edu

October 24, 2018

**Abstract**

We show that a wide variety of non-linear cellular automata (CAs) can be decomposed into a *quasidirect product* of linear ones. These CAs can be predicted by parallel circuits of depth $\mathcal{O}(\log^2 t)$ using gates with binary inputs, or $\mathcal{O}(\log t)$ depth if "sum mod $p$" gates with an unbounded number of inputs are allowed. Thus these CAs can be predicted by (idealized) parallel computers much faster than by explicit simulation, even though they are non-linear.

This class includes any CA whose rule, when written as an algebra, is a solvable group. We also show that CAs based on nilpotent groups can be predicted in depth $\mathcal{O}(\log t)$ or $\mathcal{O}(1)$ by circuits with binary or "sum mod $p$" gates respectively.

We use these techniques to give an efficient algorithm for a CA rule which, like elementary CA rule 18, has diffusing defects that annihilate in pairs. This can be used to predict the motion of defects in rule 18 in $\mathcal{O}(\log^2 t)$ parallel time.

PACS Keywords: 02.10, 02.70, 05.45, 46.10

## 1   Introduction

The *direct product* is a very basic notion in mathematics. Two algebras, dynamical systems, or members of any other category can be paired so that their components act independently of each other: in algebras, $(a_1, b_1) \cdot (a_2, b_2) = (a_1 \cdot a_2, b_1 \cdot b_2)$.

More generally, we can have a *quasidirect product* where the first component is independent of the second, but not vice versa:

$$(a_1, b_1) \cdot (a_2, b_2) = (a_1 \cdot a_2, \; b_1 \odot_{a_1,a_2} b_2)$$

where $a_1$ and $a_2$ determine some *local operation* $\odot_{a_1,a_2}$ on the $b$'s. We will be particularly interested in the case where the $b$'s form an Abelian group, and the local operations have the form

$$b_1 \odot_{a_1,a_2} b_2 = f_{a_1,a_2}(b_1) + g_{a_1,a_2}(b_2) + h \tag{1}$$

We will denote such a product by $A \otimes B$.

The *semidirect product* is a special case from group theory, in which $g$ is the identity and $f$ depends only on $a_2$:

$$(a_1, b_1) \cdot (a_2, b_2) = (a_1 \cdot a_2, \; f_{a_2}(b_1) \cdot b_2)$$

For instance, if $B$ is a *normal subgroup* of a group $G$ (so that $g^{-1}bg \in B$ for all $g \in G$ and $b \in B$) and if $A$ is a *complement* of $B$ (so that every $g \in G$ can be written uniquely as $g = ab$ where $a \in A$ and $b \in B$) then

$$g_1 \cdot g_2 = a_1 b_1 \cdot a_2 b_2 = a_1 a_2 \cdot (a_2^{-1} b_1 a_2) b_2$$

and $G$ is a semidirect product $A \otimes B$ with $f_a(b) = a^{-1}ba$.

This idea can be extended to dynamical systems. Rather than a direct product where two components evolve independently as in $(a', b') = \Phi(a, b) = (f(a), g(b))$, we can have a quasidirect product of the form

$$\Phi(a, b) = (f(a), g_a(b))$$

The second component becomes a non-autonomous dynamical system, varying in time and space in a way controlled by the first component. If a dynamical system can be decomposed in this way, and if we have efficient algorithms to predict both $f$ and the non-autonomous $g$, then we can predict the system as a whole.

Cellular automata (CAs) are dynamical systems on the space of sequences over some finite alphabet, of the form

$$\Phi(a)_i = \phi(a_{i-r}, \ldots, a_i, \ldots, a_{i+r})$$

where $r$ is the *radius* of the rule. By combining blocks of $2r$ sites together, as shown in figure 1, we can convert any CA into one with $r' = 1/2$ where each site has only two predecessors in a staggered space-time:

$$\Phi(a)_i = \phi(a_{i-1/2}, a_{i+1/2})$$

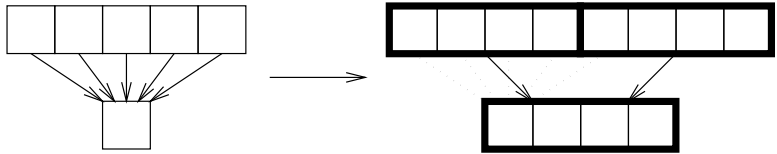

Figure 1: By blocking together $2r$ sites, we can transform any CA into one on a staggered space-time with $r' = 1/2$. Here $r = 2$.

(See [19] for a study of this transformation's algebraic properties.) We can then think of the CA rule as a binary algebra, where $\phi(a,b) = a \cdot b$ or $ab$ for short. The light-cone below an initial row becomes

$$
\begin{array}{cccc}
a_0 & a_1 & a_2 & a_3 \\
a_0a_1 & a_1a_2 & a_2a_3 \\
(a_0a_1)(a_1a_2) & (a_1a_2)(a_2a_3) \\
\big((a_0a_1)(a_1a_2)\big)\big((a_1a_2)(a_2a_3)\big)
\end{array}
$$

and so on. Several authors have used this approach to explore CA properties such as partial reversibility [9] and periodicity [24].

Predicting a cellular automaton $t$ time-steps into the future is believed to be no easier in general than simulating it explicitly. To do this, we have to calculate all the CA states in a light-cone of depth $t$, which takes $\mathcal{O}(t^2)$ serial computation steps ($\mathcal{O}(t^{d+1})$ in $d$ dimensions) or $\mathcal{O}(t)$ in parallel.

However, in [17] we show that CAs whose rules satisfy various algebraic identities can be predicted in parallel time $\mathcal{O}(\log^k t)$ for some $k$, qualitatively faster than explicit simulation. We term these CAs *quasi-linear:* they are non-linear, but efficiently predictable nonetheless.

It would be highly surprising if this were true for all CAs: since CAs exist which can simulate universal Turing machines (e.g. [14]), predicting a CA for a linear (or polynomial) amount of time is **P**-complete in general [11]. Particular CA rules based on the Ising model or on majority-voting in three or more dimensions can also be shown to be **P**-complete [18].

Therefore, it seems worthwhile to extend the class of quasi-linear CAs as far as possible, to explore what is probably a very rich hierarchy between linear dynamical systems and computationally universal ones.

A preliminary version of these results appeared in [20].

## 2 Definitions

An *algebra* $(A, \cdot)$ is a function from $A \times A$ to $A$, written $a \cdot b$ or simply $ab$. The *order* of an algebra is the number of elements in $A$. We will concern ourselves here with algebras of finite order.

The *direct product* $A \times B$ of two algebras is the set of pairs $(a, b)$, with $(a_1, b_1)(a_2, b_2) = (a_1 a_2, b_1 b_2)$.

A *quasigroup* is an algebra whose multiplication table is a *Latin square*, in which every element of $A$ occurs once in each row and each column. Quasigroups correspond to *permutive* CAs, in which $\phi(a, b)$ is a one-to-one function of each of its inputs (more generally, its leftmost and rightmost inputs) when the others are held fixed.

An *identity* is an element 1 such that $1a = a1 = a$ for all $a$. An *inverse* of an element $a$ is an element $a^{-1}$ such that $a^{-1}a = aa^{-1} = 1$.

An algebra is *associative* if $a(bc) = (ab)c$ for all $a, b, c \in A$. An associative algebra is called a *semigroup*. An associative quasigroup is a *group*. Groups have identities and inverses.

An algebra is *commutative* if $ab = ba$ for all $a, b \in A$. Commutative groups are called *Abelian*. The *cyclic group* $\mathbb{Z}_p = \{0, 1, \ldots, p-1\}$ with addition mod $p$ is Abelian.

A function $f$ on an algebra is a *homomorphism* if $f(ab) = f(a)f(b)$. Homomorphisms of Abelian groups can be represented as matrices. An *isomorphism* is a one-to-one and onto homomorphism; we will write $A \cong B$ if $A$ and $B$ are isomorphic. An *automorphism* is an isomorphism from an algebra to itself.

A *subgroup* $B$ of a group $A$ is a subset such that $b_1 b_2 \in B$ for all $b_1, b_2 \in B$. The subgroup generated by all possible products of elements in a subset $S$ is written $\langle S \rangle$.

A subgroup $B$ is *normal* if $a^{-1}ba \in B$ for all $b \in B$ and all $a \in A$. For any normal subgroup $B$ of $A$, there is a *factor group* $A/B$ and a homomorphism from $A$ to $A/B$ that sends all elements of $B$ to 1.

In a non-Abelian group, the *commutator* of $a$ and $b$ is $[a, b] = a^{-1}b^{-1}ab$, so $ab = ba[a, b]$. The *commutator subgroup* $G' = \langle [G, G] \rangle$ of a group $G$ is the set of all elements that can be written as products of commutators; then $G/G'$ is Abelian.

As a model of computation, we will consider families of circuits of varying depth and with different types of gates. The following classes of problems are those for which for all $n$, there is a circuit $\mathcal{C}_n$ of depth $\mathcal{O}(\log^k n)$ and size polynomial in $n$ that outputs the answer for inputs of size $n$:

- $\mathbf{NC}^k$ if the gates are ANDs and ORs with binary inputs,

- $\mathbf{AC}^k$ if the gates are ANDs and ORs with an unbounded number of inputs (unbounded fan-in), and

- $\mathbf{ACC}^k[p]$ if the gates are ANDs, ORs, and "sum mod $p$" with an unbounded number of inputs. $\mathbf{ACC}^k$ is the union $\cup_p \mathbf{ACC}^k[p]$.

Then $\mathbf{NC} = \cup_k \mathbf{NC}^k = \cup_k \mathbf{AC}^k$ is the class of problems solvable in polylogarithmic time by an idealized parallel computer with a polynomial number of processors, and is considered a good definition of problems that are efficiently parallelizable. (A more realistic model of parallel computation would take the time and cost of communication between processors into account, assuming some finite-dimensional topology.)

Families of circuits are *uniform* if there is a simple algorithm that generates $\mathcal{C}_n$ when given $n$ as input. Typically **LOGSPACE**-uniformity is used, i.e. a Turing machine that generates $\mathcal{C}_n$ using $\mathcal{O}(\log n)$ bits of memory.

**P** is the class solvable by a deterministic Turing machine in polynomial time. It is easy to see that $\mathbf{NC} \subseteq \mathbf{P}$, but like $\mathbf{P} \subseteq \mathbf{NP}$ this inclusion is believed, but not known, to be proper. From the definitions we have

$$\mathbf{AC}^0 \subset \mathbf{ACC}^0[2] \subset \mathbf{ACC}^0 \subseteq \mathbf{NC}^1 \subseteq \mathbf{AC}^1 \subseteq \mathbf{ACC}^1 \subseteq \mathbf{NC}^2 \subseteq \cdots \subseteq \mathbf{NC} \subseteq \mathbf{P}$$

The parity function is clearly in $\mathbf{ACC}^0[2]$, and Ajtai and Furst et al. [1, 10] have shown that it is not in $\mathbf{AC}^0$. Razborov [25] has shown that majority is in $\mathbf{NC}^1$ but not in $\mathbf{ACC}^0[2]$, and Smolensky [28] has shown that $\mathbf{ACC}^0[p]$ and $\mathbf{ACC}^0[q]$ are incomparable if $p$ and $q$ are distinct primes. Thus the first and second inclusions are proper. However, for all anyone has been able to prove, $\mathbf{ACC}^0[6]$ could be equal to **P** (or even **NP** for that matter).

A problem is **P**-*complete* if instances of any other problem in **P** can be converted to it by a **LOGSPACE** algorithm. It is generally believed that **P**-complete problems are *inherently sequential*, and cannot be efficiently parallelized; if any **P**-complete problem is in **NC**, then $\mathbf{P} = \mathbf{NC}$ and all polynomial-time problems can be solved in polylogarithmic parallel time [11, 23].

The canonical **P**-complete problem is CIRCUIT VALUE: what is the output of a given Boolean circuit, given the truth values of its inputs? Since truth values at each level of the circuit can affect those on the next level in arbitrary ways, it is hard to see how to get the output without going through the circuit level-by-level.

An *algebraic circuit* over an algebra $(A, \cdot)$ is a circuit where each gate outputs the product $a \cdot b$ of its inputs, rather than implementing the standard Boolean functions; the CIRCUIT VALUE problem for various classes of algebras has been studied by a number of authors [2, 3, 22]. Predicting an $r = 1/2$ CA is clearly a special case of CIRCUIT VALUE, where the circuit has a simple periodic structure in space and time. However, a number of the results we prove below for CAs will in fact hold for algebraic circuits of arbitrary shape.

As shorthand, we will say a CA is in $\mathbf{NC}^k$ ($\mathbf{AC}^k$, $\mathbf{ACC}^k$) if it can be predicted by circuits in these classes. Since a CA's input consists of $n = 2rt+1 = \mathcal{O}(t)$ initial sites, we will use $n$ and $t$ interchangeably in our $\mathcal{O}$'s.

## 3 Non-autonomous additive CAs

We will repeatedly use the fact that

**Lemma 0.** *The sum of $n$ elements of a finite Abelian group can be calculated in constant depth with* **ACC** *gates, or in $\mathcal{O}(\log n)$ depth with binary gates.*

**Proof.** Any finite Abelian group can be written as a direct product of cyclic groups $G = \mathbb{Z}_{p_1} \times \mathbb{Z}_{p_2} \times \cdots \times \mathbb{Z}_{p_k}$. Addition of $n$ elements of $\mathbb{Z}_{p_i}$ can be carried out with a single **ACC**$[p_i]$ gate. Since $a \bmod p_i = (aq \bmod p_i q)/q$, **ACC**$[p_i]$ gates can be simulated by **ACC**$[p_i q]$ gates for any $q$, and in particular by **ACC**$[p]$ gates where $p = \mathrm{lcm}(p_1, p_2, \ldots, p_k)$. Alternately, a tree of binary gates of depth $\mathcal{O}(\log n)$ can add the elements in pairs. ∎

Now if a CA's algebra is an Abelian group $\phi(a, b) = a + b$, a simple Green's function method with Pascal's triangle coefficients [17, 26] allows us to predict the CA in $\mathcal{O}(t)$ serial time or $\mathcal{O}(\log t)$ parallel time, i.e. in **NC**$^1$. In fact, this algorithm is in **LOGSPACE**-uniform **ACC**$^0$ since we can generate the $t$'th row of Pascal's triangle using $\mathcal{O}(\log t)$ space.

More generally, if a CA is of the form

$$\phi(a_0, a_1, \ldots, a_{2r}) = f_0 a_0 + f_1 a_1 + \cdots + f_{2r} a_{2r} + h \tag{2}$$

where the $f_i$ are matrix-valued homomorphisms of an Abelian group $(A, +)$ and $h$ is a constant element of $A$, we can represent the CA rule as a polynomial [16]

$$G(x) = f_0 + f_1 x + f_2 x^2 + \cdots + f_{2r} x^{2r}$$

plus the constant $h$. Then we can think of the coefficients of $G^t(x)$ as the $t$'th row of a Green's function for the CA, and for each $t$ the final state is

$$s = \sum_{i=0}^{t} G_i^t a_i + h \sum_{t'=0}^{t-1} \sum_{i=0}^{t'} G_i^{t'}$$

The multiplications can be carried out in constant depth, and by lemma 0 the sum is in **ACC**$^0$. So any such CA is in **ACC**$^0$.

We now show that a non-autonomous version of (2) is still efficiently predictable in parallel.

**Lemma 1.** *A non-autonomous CA of the form (2), where the $f_i$ are homomorphisms of an Abelian group $(A, +)$ and $h$ is an element of $A$, which vary in space and time independently of the CA state, is in* **ACC**$^1$.

**Proof.** For simplicity, we will prove this for $r = 1/2$, where the CA rule is an operation of the form (1). A larger radius will simply increase the width of the light-cones, and the computation time, by a constant.

Call the states in the light-cone $s_{t,x}$ with $t = 0$ at the initial row, and $s_{t,0}$ the leftmost state in the $t$'th row, as shown in figure 2. Then $s_{t,x}$'s predecessors are $s_{t-1,x}$ and $s_{t-1,x+1}$, and we write

$$s_{t,x} = \phi(s_{t-1,x}, s_{t-1,x+1}) = f_{t,x}(s_{t-1,x}) + g_{t,x}(s_{t-1,x+1}) + h_{t,x}$$

(where we have replaced $f_1$ and $f_2$ with $f$ and $g$ for clarity) where $f_{t,x}$ and $g_{t,x}$ are homomorphisms of $(A, +)$ and $h_{t,x}$ is an element of $A$, all of which vary with $t$ and $x$.

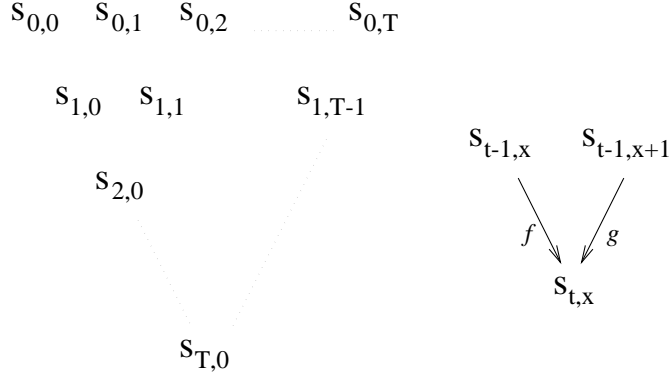

Figure 2: The labelling scheme used in the text for the light-cone below the initial row.

Now if $(t', x')$ is in the light-cone below $(t, x)$, define $C_{t',x'|t,x}$ as the coefficient of $s_{t,x}$ in $s_{t',x'}$. Then we have

$$C_{t',x'|t,x} = \begin{cases} 1 & \text{if } t' = t \text{ and } x' = x \\ \left(f_{t',x'} \circ C_{t'-1,x'|t,x}\right) + \left(g_{t',x'} \circ C_{t'-1,x'+1|t,x}\right) & \text{if } t' > t \end{cases}$$

Then it is straightforward to show inductively that the state $s_{T,0}$ at the bottom of a light-cone $T$ steps high, with initial row $s_{0,x}$ for $0 \leq x \leq T$, is

$$s_{T,0} = \sum_{x=0}^{T} C_{T,0|0,x}(s_{0,x}) + \sum_{t=1}^{T} \sum_{x=0}^{T-t} C_{T,0|t,x}(h_{t,x}) \qquad (3)$$

Since we are given the $h_{t,x}$, we can add all $(T+1)(T+2)/2$ of these terms together in depth $\mathcal{O}(\log T)$ with binary gates or in constant depth with an **ACC** circuit by lemma 0; but we have to calculate the $C_{T,0|t,x}$ first.

We will calculate the $C_{t',x'|t,x}$ using a divide-and-conquer strategy. The influence of $s_{t,x}$ on $s_{t',x'}$ has to go through the intervening sites, so for any $t''$ where $t < t'' < t'$, we can write

$$C_{t',x'|t,x} = \sum_{x''=\max(x-t''+t,x')}^{\min(x,x'+t-t'')} C_{t',x'|t'',x''} \circ C_{t'',x''|t,x} \qquad (4)$$

as shown in figure 3.

We can then use induction on increasing time intervals. Assume that the $C_{t',x'|t'',x''}$ and $C_{t'',x''|t,x}$ are known. Since the sum in (4) has at most $(t'-t)/2+1$ terms and the compositions can be done simultaneously in constant time, by lemma 0 we can calculate $C_{t',x'|t,x}$ in depth $\mathcal{O}(\log(t'-t))$ with binary gates or constant depth with **ACC** gates.

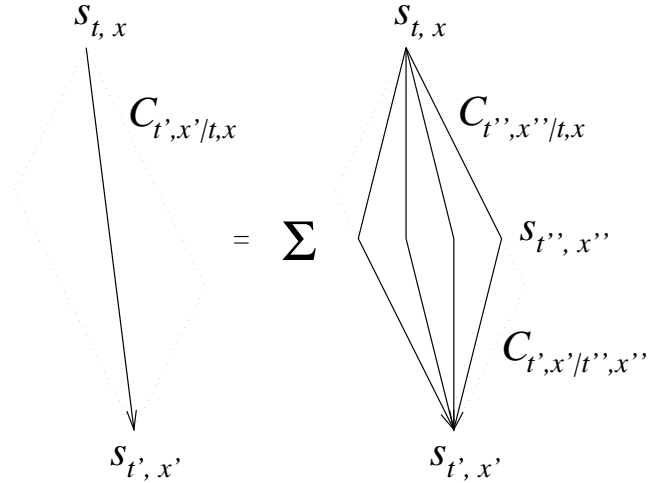

Figure 3: The divide-and-conquer strategy of equation 3.

In particular, let $t'' = \lfloor (t'-t)/2 \rfloor$ so that we double the time interval at each stage. Start by calculating (in parallel) all the $C_{t',x'|t,x}$ where $t'-t=1$, then all those with $t'-t=2$, and so on until we reach $t'-t=T$. This takes $\mathcal{O}(\log T)$ stages of induction, giving a total depth $\mathcal{O}(\log^2 T)$ with binary gates or $\mathcal{O}(\log T)$ with **ACC** gates, i.e. $\mathbf{NC}^2$ or $\mathbf{ACC}^1$.

Once we have $C_{T,0|t,x}$ for all $t,x$ we can easily add the sum in (3) as stated above to get the final state $s_{T,0}$. So predicting the CA can be done in $\mathbf{ACC}^1$.
∎

We can generalize this further in two ways. First, for these sums to work, $(A,+)$ simply needs to be commutative and associative, i.e. it can be a commutative semigroup rather than a group. Secondly, this algorithm works in any number of dimensions; the number of sites at $t''$ between $t$ and $t'$ is proportional to $(t'-t)^d$, so the divide-and-conquer algorithm works just as fast. So in full generality we can state the following theorem:

**Theorem 1.** *Any complexity class of CAs containing $\mathbf{ACC}^1$ is closed under quasidirect products with commutative semigroups where the local operations are of the form (1) where $f$ and $g$ are homomorphisms, or $(2r+1)$-ary operations of the form (2) where the $f_i$ are homomorphisms. Therefore, in any number of dimensions, suppose a CA is an iterated quasidirect product of the form*

$$C = (((C_0 \otimes C_1) \otimes C_2)\cdots) \otimes C_n$$

*where $C_0$ is in $\mathbf{ACC}^1$ and the $C_i$ for $i > 0$ are non-autonomous additive CAs of the form*

$$\phi_i(a_1, a_2, \ldots, a_k) = f_{i,1}(a_1) + f_{i,2}(a_2) + \cdots + f_{i,k}(a_k) + h_i$$

where the $f_{i,j}$ are homomorphisms of a commutative semigroup $(A_i, +)$ and the $h_i$ are elements of $A_i$, depending on the state of $C_{i-1}$ at each point in space-time. Then $C$ is in $\mathbf{ACC}^1$.

**Proof.** Since $C_0$ is in $\mathbf{ACC}^1$, we can calculate its state everywhere in the light-cone simultaneously with a circuit $t^2$ times as large. This and Lemma 1 tells us how to calculate $C_1$ everywhere in the light-cone, from which we can calculate $C_2$ everywhere, and so on. Iterate $n$ times. ∎

**Example 1.** The homomorphisms of $\mathbb{Z}_2$ are zero and the identity, so the functions expressible as $f(a) + g(b) + h$ are $0, 1, a, \overline{a}, b, \overline{b}, a \oplus b$ and $\overline{a \oplus b}$. So any quasidirect product $C \otimes \mathbb{Z}_2$ where the states of $C$ select among these eight functions on $\mathbb{Z}_2$ is in $\mathbf{ACC}^1$ if $C$ is. In particular, six quasigroups of order 4, including $\mathbb{Z}_2^2$, $\mathbb{Z}_4$, and four non-associative ones, can be written as quasidirect products $\mathbb{Z}_2 \otimes \mathbb{Z}_2$.

**Example 2.** The automorphisms of $\mathbb{Z}_3$ are $f(a) = \pm a$. All quasigroups of 3 elements can be expressed as $\pm a \pm b + h$. Therefore, any permutive CA which is a quasidirect product $C \otimes A$, where $A$ has three elements, is in $\mathbf{ACC}^1$ if $C$ is.

**Example 3.** Suppose a CA has a vector-valued state $(a_1, \ldots, a_k)$ at each site. Any $r = 1/2$ CA of the form

$$\phi((a_1, \ldots, a_k), (b_1, \ldots, b_k)) = \\ (P_1(a_1, b_1), P_2(a_1, a_2, b_1, b_2), \ldots, P_k(a_1, \ldots, a_k, b_1, \ldots, b_k))$$

(or the obvious generalization for larger $r$) is in $\mathbf{ACC}^1$ if each $P_i$ is linear in $a_i$ and $b_i$ but has arbitrary dependence on $a_j$ and $b_j$ for $j < i$.

Since $\log(n^c) = \mathcal{O}(\log n)$. the divide-and-conquer algorithm in lemma 1 actually reduces any polynomial size circuit of polynomial depth $n^c$ to an $\mathbf{ACC}^1$ circuit as long as the gates are functions of the form (2). Thus we can state the following:

**Corollary.** *The* CIRCUIT VALUE *problem for algebraic circuits, over an algebra formed by an iterated quasidirect product of commutative semigroups where the local operations are of the form (2), is in $\mathbf{ACC}^1$.*

## 4  CAs based on solvable groups

One interesting class of CAs consists of those for which $\phi(a, b) = a \cdot b$ is a non-Abelian group. Because of their non-commutativity these CAs are non-linear, i.e. they do not obey a principle of superposition, so Green's function techniques don't work. In [17] we show an $\mathcal{O}(\log t)$ algorithm for one such group, the Quaternions $Q_8$, but other non-Abelian groups such as $S_3$, the group of permutations of three elements, are left as open problems.

We now show that a large class of finite groups have CAs in $\mathbf{ACC}^1$. First:

**Lemma 2.** *The set of algebras whose CAs can be predicted in a given amount of serial or parallel time (up to a multiplicative constant) is closed under finite direct products, subgroups, and homomorphisms.*

**Proof.** For finite direct products $G_1 \times G_2 \times \cdots$, simply predict each of the $G_i$, either sequentially or in parallel. For subgroups, clearly an algorithm that predicts an algebra also predicts any of its subgroups. For a homomorphic image $H = f(G)$, choose a pre-image $f^{-1}(h)$ for each element $h \in H$ in the initial conditions, use the algorithm for $G$, and then apply $f$ to return to $H$. Since $f$ is a homomorphism, a simple induction shows that $H^t(h_0, \ldots, h_t) = f(G^t(f^{-1}(h_0), \ldots, f^{-1}(h_t)))$ for all $t$. ∎

In the language of universal algebra, this makes the set of CAs in a given complexity class a *pseudovariety* [8]. It would be wonderful if the pseudovariety corresponding to $\mathbf{ACC}^1$ or $\mathbf{NC}$, say, were *finitely presented*: that is, if a finite set of algebraic identities were necessary and sufficient for a CA to be in a given parallel complexity class.

Now recall the following definition from group theory.

**Definition [12].** The *derived series* of a group $G$ is the series of normal subgroups $G = G_0 \supseteq G_1 \supseteq \cdots$ where $G_{i+1} = \langle [G_i, G_i] \rangle = G'_i$ is the commutator subgroup of $G_i$. A group is *solvable* if its derived series ends in the identity $\{1\}$ after a finite number of steps. Each of the factors $G_i/G_{i+1}$ is Abelian.

For instance, the derived series of $S_4$, the group of permutations of 4 objects, is $S_4 \supset A_4 \supset \mathbb{Z}_2^2 \supset \{1\}$ where $A_4$ is the set of even permutations and $\mathbb{Z}_2^2$ consists of the identity and the permutations $(12)(34)$, $(13)(24)$ and $(14)(23)$.

**Definition.** A group is *polyabelian* if it can be written as a semidirect product of Abelian groups $((A_1 \otimes A_2) \otimes \cdots) \otimes A_k$ with the parentheses associated to the left. (This could also be called an *iterated split extension*.)

Clearly a CA based on a polyabelian group is in $\mathbf{ACC}^1$ by theorem 1. Then:

**Theorem 2.** *Any CA whose algebra is a solvable group is in $\mathbf{ACC}^1$.*

**Proof.** We will show that any solvable group is isomorphic to a subgroup of a polyabelian group. Recall [29] that the *wreath product* $A \wr B$ is a semidirect product $B \otimes A^B$ where $A^B$ is the set of functions $\alpha$ from $B$ to $A$ and elements of $B$ permute their components. In other words,

$$(b_1, \alpha_1)(b_2, \alpha_2) = (b_1 b_2, f_{b_2}(\alpha_1) \circ \alpha_2) \text{ where } f_b(\alpha)(x) = \alpha(bx)$$

The wreath product is useful for the following reason: if $G$ has a normal subgroup $N$, then $G$ is isomorphic to a subgroup of $N \wr (G/N)$.

Since every solvable group has an Abelian normal subgroup (the last non-trivial group in its derived series, whose commutator subgroup is $\{1\}$), by induction it can be embedded in a wreath product of Abelian groups. In particular, if $G = G_0 \supset G_1 \supset \cdots \supset G_k = \{1\}$, then

$$G \subset H_k \wr (H_{k-1} \wr \cdots (H_2 \wr H_1))$$

where $H_k = G_{k-1}/G_k$ are the Abelian factor groups. This is a semidirect product of Abelian groups

$$G \subset ((K_1 \otimes K_2) \cdots \otimes K_{k-1}) \otimes K_k$$

where $K_1 = H_1$ and $K_{i+1} = H_i^{K_i}$. Since $G$ can be embedded in a polyabelian group, by lemma 2 a CA with $G$ as its algebra is in $\mathbf{ACC}^1$. ∎

As before, we can state the corollary for circuits in general:

**Corollary.** *The* CIRCUIT VALUE *problem for algebraic circuits over solvable groups is in* $\mathbf{ACC}^1$.

For instance, the Quaternions $Q_8$ have the derived series $Q_8 \supset \{\pm 1\} \supset \{1\}$ since $[a, b] = \pm 1$ for any $a, b$ and $[\pm 1, \pm 1] = 1$. The factor groups are $H_1 = Q_8/\{\pm 1\} \cong \mathbb{Z}_2^2$ and $H_2 = \{\pm 1\} \cong \mathbb{Z}_2$, so $Q_8$ is a subgroup of $\mathbb{Z}_2 \wr \mathbb{Z}_2^2$. This is a semidirect product $\mathbb{Z}_2^2 \otimes \mathbb{Z}_2^4$ since $\mathbb{Z}_2^{\mathbb{Z}_2^2} \cong \mathbb{Z}_2^4$.

In general, the $K_i$ grow alarmingly in size, like $\underbrace{n^{n^{\cdot^{\cdot^{\cdot^n}}}}}_{i \text{ times}}$. But most groups don't require embedding in such large wreath products. Many small groups are themselves polyabelian, including the dihedral groups, groups of order $p^3$ for $p$ an odd prime, any group of square-free order, all groups of order $p^2q$ where $p$ and $q$ are primes, and so on [4, 12]. All groups of order less than 32 are polyabelian except the *dicyclic* or *generalized quaternion* groups, which are factor groups of polyabelian groups twice their size, and the *binary tetrahedral group* of order 24, which is a subgroup of a polyabelian group of order $12288 = 2^{12} \cdot 3$ [27].

The smallest non-solvable group is $A_5$, the simple group of order 60 (also called the icosahedral group). Since polyabelian groups are solvable (if $G = ((A_0 \otimes A_1) \otimes A_2) \otimes \ldots$, then $G' \subseteq (A_1 \otimes A_2) \otimes \ldots$) and since subgroups and factors of solvable groups are also solvable [12], this group's CA cannot be predicted by these methods.

This leaves us with the following open question: is there an algorithm in $\mathbf{ACC}^1$, or $\mathbf{NC}^k$ for some $k$, for predicting CAs based on arbitrary finite groups? Barrington [2] and Beaudry et al. [3] have provided strong evidence to the contrary: since multiplication in $A_5$ can simulate NAND gates, circuits with non-solvable gates have $\mathbf{P}$-complete CIRCUIT VALUE problems. Unless there is some way to exploit the periodic structure of the circuit corresponding to a CA's evolution, then, we would have

**Conjecture.** *CAs based on non-solvable groups such as $A_5$ are* $\mathbf{P}$-*complete.*

Finally, we note that since the divide-and-conquer algorithm of lemma 1 applies to circuits of arbitrary shape, we have actually shown that CIRCUIT VALUE is in $\mathbf{ACC}^1$ for solvable groups, or for polyabelian algebras formed inductively as in theorem 1. This is somewhat more general than the result of Beaudry et al. [3], who show that CIRCUIT VALUE is in the class $\mathbf{DET}$ for solvable semigroups, which is not known to be comparable with $\mathbf{ACC}^1$.

## 5  CAs based on nilpotent groups

We now show that a subset of the solvable groups have CAs in $\mathbf{ACC}^0$.

**Definition [12].** The *lower central series* of a group $G$ is the series of normal subgroups $G = \Gamma_1 \supseteq \Gamma_2 \supseteq \cdots$ where $\Gamma_{i+1} = \langle[\Gamma_i, G]\rangle$. In other words, $\Gamma_2$ is the commutator subgroup $G'$, $\Gamma_3$ is the subgroup generated by 3-element commutators $[[a,b],c]$, and so on. If $\Gamma_{k+1} = \{1\}$ for some $k$, we say that $G$ is *nilpotent of class $k$*, and all commutators with more than $k$ elements are 1. The nilpotent groups form a proper subset of the set of solvable groups.

For instance, a nilpotent group of class 1 is simply an Abelian group. A nilpotent group of class 2 has commutators which commute with everything, so that $[[a,b],c] = 1$ (such as the Quaternions where $[a,b] = \pm 1$ and $[\pm 1, c] = 1$). And so on.

**Theorem 3.** *Any CA whose rule is a nilpotent group is in* **LOGSPACE**-*uniform* $\mathbf{ACC}^0$.

**Proof.** First, consider a nilpotent group of class 2. If its initial conditions are $a_0, a_1, \ldots, a_t$, the leftmost two columns of its light-cone are

$$
\begin{array}{cc}
a_0 & a_1 \\
a_0 a_1 & a_1 a_2 \\
a_0 a_1^2 a_2 & a_1 a_2^2 a_3 \\
a_0 a_1^3 a_2^3 a_3 \cdot [a_1, a_2]^{-1} & a_1 a_2^3 a_3^3 a_4 \cdot [a_2, a_3]^{-1}
\end{array}
$$

The commutator arises since $a_2 a_1 = a_1 a_2 [a_1, a_2]^{-1}$, so

$$(a_0 a_1^2 a_2)(a_1 a_2^2 a_3) = a_0 a_1^3 a_2 [a_1, a_2]^{-1} a_2^2 a_3$$

Since commutators commute with everything we can move it to the right, leaving the $a_i$ in sorted order. Continuing in this way we get an "Abelian part" $\prod_{i=0}^{t} a_{x+i}^{\binom{t}{i}}$, times powers of commutators $[a_i, a_j]$ where $i < j$. For $t = 4, 5, 6$ these commutators are

$$[a_1, a_2]^{-4} [a_2, a_3]^{-4} \cdot [a_1, a_3]^{-1}$$

$$[a_1, a_2]^{-10} [a_2, a_3]^{-24} [a_3, a_4]^{-10} \cdot [a_1, a_3]^{-5} [a_2, a_4]^{-5} \cdot [a_1, a_4]^{-1}$$

$$[a_1, a_2]^{-20} [a_2, a_3]^{-84} [a_3, a_4]^{-84} [a_4, a_5]^{-20} \cdot$$
$$[a_1, a_3]^{-15} [a_2, a_4]^{-35} [a_3, a_5]^{-15} \cdot [a_1, a_4]^{-6} [a_2, a_5]^{-6} \cdot [a_1, a_5]^{-1}$$

Since each site $s_{t,x}$ in a light-cone is a product of $s_{t-1,x}$, which contains $a_j^{\binom{t-1}{j-x}}$, and $s_{t-1,x+1}$, which contains $a_i^{\binom{t-1}{i-x-1}}$, and since $a_j^q a_i^p = a_i^p a_j^q [a_i, a_j]^{-pq}$, we get $\binom{t-1}{j-x}\binom{t-1}{i-x-1}$ factors of $[a_i, a_j]^{-1}$ in $s_{t,x}$ when we pass these powers of $a_i$ and $a_j$ through each other. Each of these commutators comes down to the final site $s_{T,0}$ in $\binom{T-t}{x}$ ways, so

$$s_{T,0} = \prod_{0 \leq i \leq T} a_i^{\binom{T}{i}} \prod_{0 < i < j < T} [a_i, a_j]^{-\sum_{t=0}^{T} \sum_{x=\max(0, j-t-1)}^{\min(T-t, i-1)} \binom{T-t}{x}\binom{t-1}{j-x}\binom{t-1}{i-x-1}}$$

(If $[a,b]^2 = 1$ for all $a, b$, this is just the algorithm for $Q_8$ given in [17].)

In general, if we have a nilpotent group of class $k$, then the final site can always be written

$$s_{T,0} = \prod_i a_i^{c_i^{(1)}} \prod_{i,j} [a_i, a_j]^{c_{ij}^{(2)}} \cdots \prod_{i_1, i_2, \ldots, i_k} [[\cdots [a_{i_1}, a_{i_2}], \cdots], a_{i_k}]^{c_{i_1 i_2 \cdots i_k}^{(k)}} \qquad (5)$$

for some set of exponents $c_{i_1 i_2 \ldots i_j}^{(j)}$ ($1 \leq j \leq k$). This is clear by induction: by sorting the $a_i$ on the left, we generate 2-element commutators $[a_i, a_j]$. By sorting these we generate 3- and 4-element commutators, and so on, until we reach $k$-element commutators which commute with everything.

For each $t$, then, we have an expression (5) with $\mathcal{O}(t^k)$ terms. Evaluating an expression of length $m$ in a solvable group can be done in depth $\mathcal{O}(\log m)$ with binary gates or constant depth with **ACC** gates [2]. So nilpotent CAs are in $\mathbf{ACC}^0$.

Finally, we show that this family of circuits is **LOGSPACE**-uniform, i.e. the circuit for predicting the CA $t$ time-steps into the future can be generated by an algorithm using only $\mathcal{O}(\log t)$ bits of memory. This is equivalent to calculating all the $c_{i_1 i_2 \ldots i_j}^{(j)}$. Since $j$-element commutators are made by crossing commutators with fewer elements, each $c^{(j)}$ at $s_{t,x}$ is a polynomial of $c^{(j')}$'s at $s_{t-1,x}$ and $s_{t-1,x+1}$ where $j' < j$. These are polynomials of smaller commutators, and so on.

So to get each $c^{(j)}$, we have to calculate a tree of smaller $c^{(j')}$. This tree is of constant depth (at most $k$), and it ends in leaves of the form $c_i^{(1)} = \binom{t}{i}$ which can be calculated in $\mathcal{O}(\log t)$ space. So each of the $c^{(j)}$ can be calculated in logarithmic space, and since there are only polynomially many of them (which can be indexed with $\mathcal{O}(\log t)$ bits) all of them can. ∎

Since the same technique shows allows us to calculate the $c^{(j)}$ for any algebraic circuit over a nilpotent group, we have the following corollary:

**Corollary.** *Any algebraic circuit over a nilpotent group is equivalent to an $\mathbf{ACC}^0$ circuit.*

Note that this is not the same as saying that CIRCUIT VALUE in general for circuits over nilpotent groups is in $\mathbf{ACC}^0$; for this to be true, we would have to be able to take any circuit of arbitrary shape, and convert its topology, in constant parallel time, into the $c^{(j)}$'s. We can say rather that any single circuit over a nilpotent group can be converted to an $\mathbf{ACC}^0$ function of its inputs, and so e.g. **LOGSPACE**-uniform families of algebraic circuits become **LOGSPACE**-uniform families of $\mathbf{ACC}^0$ functions. Then CIRCUIT VALUE restricted to such a family is in $\mathbf{ACC}^0$, as CA prediction is.

# 6 Defining rules with words

As another extension, suppose that we define a CA rule not simply as the product $\phi(a,b) = ab$ in a given algebra $A$, but as the value of some word such as $\phi(a,b) = a^2ba^{-1}$. CAs with larger neighborhoods or in higher dimensions can also be defined this way, where $\Phi(a)_i$ is the value of some word involving $a_i$ and its neighbors as variables. We can also apply arbitrary homomorphisms to the variables in any way we please. Call such a CA rule *word-defined* on the original algebra $A$.

Then since theorems 2 and 3 hold for circuits in general with solvable or nilpotent gates, and since each step of such a word-defined CA rule can be be written as a circuit with several gates of the original algebra, we can state the following:

**Corollary to Theorems 2 and 3.** *Any word-defined CA rule on a solvable or nilpotent group can be predicted in $\mathbf{ACC}^1$ or $\mathbf{ACC}^0$ respectively.*

Many interesting non-associative algebras can be written as words over non-Abelian groups; see [5] for a number of such constructions.

# 7 Diffusing defect dynamics

We close by showing that these techniques can in fact be applied to "classical" CA behaviors. Consider the following algebra on three symbols $\{0, 1, 1'\}$:

| $\star$ | 0 | 1 | $1'$ |
|---|---|---|---|
| 0 | 0 | 1 | $1'$ |
| 1 | 1 | 0 | 0 |
| $1'$ | $1'$ | 0 | 0 |

There are two subalgebras, $\{0,1\}$ and $\{0,1'\}$, which appear as *domains* in the CA's evolution: these are shown in black and gray in figure 4. Boundaries or *defects* between these domains act as diffusing particles, and annihilate when they meet in pairs.

**Theorem 4.** *The CA rule $\star$ shown in figure 4 is in $\mathbf{ACC}^1$.*

**Proof.** If we replace 0 with $(0,0)$, 1 with $(1,0)$, and $1'$ with $(1,1)$, then $\star$ becomes a quasidirect product of $\mathbb{Z}_2$ by $\mathbb{Z}_2$:

$$(a_1, b_1) \star (a_2, b_2) = (a_1 \oplus a_2, \ f_{a_1,a_2}(b_1) \oplus g_{a_1,a_2}(b_2))$$

where $f$ and $g$ are homomorphisms depending on $a_1$ and $a_2$:

$$f_{a_1,a_2}(b) = \begin{cases} b & \text{if } a_1 = 1, a_2 = 0 \\ 0 & \text{otherwise} \end{cases} \quad \text{and} \quad g_{a_1,a_2}(b) = \begin{cases} b & \text{if } a_1 = 0, a_2 = 1 \\ 0 & \text{otherwise} \end{cases}$$

Then by lemma 1 this CA is in $\mathbf{ACC}^1$. ∎

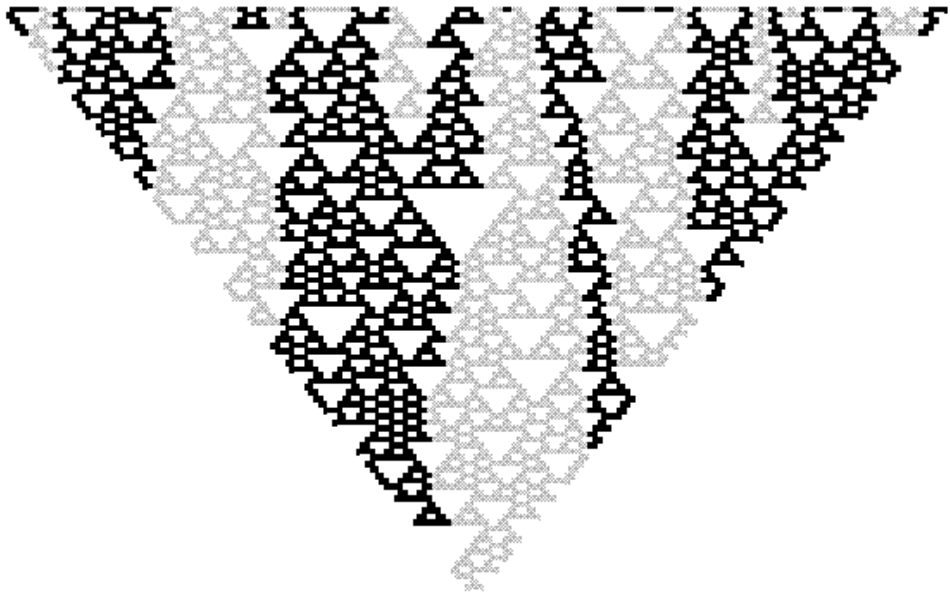

Figure 4: This CA rule has diffusing defects between domains (gray and black) that annihilate in pairs, similar to those in elementary CA rule 18. It can be predicted by $\mathbf{ACC}^1$ circuits, and it matches rule 18 exactly for the motion of a single defect.

This rule is closely related to the two-state, $r = 1$ CA rule 18 (by the numbering system of [30]):

$$\begin{array}{rcccccccc} a_{i-1}a_i a_{i+1} : & 111 & 110 & 101 & 100 & 011 & 010 & 001 & 000 \\ a'_i : & 0 & 0 & 0 & 1 & 0 & 0 & 1 & 0 \end{array}$$

Blocking pairs of sites together gives the algebra

| $\bullet$ | 00 | 01 | 10 | 11 |
|---|---|---|---|---|
| 00 | 00 | 01 | 10 | 10 |
| 01 | 01 | 00 | 00 | 00 |
| 10 | 10 | 11 | 00 | 00 |
| 11 | 01 | 00 | 00 | 00 |

The subalgebras $\{00, 01\}$ and $\{00, 10\}$ form two domains on which the CA acts like $\mathbb{Z}_2$, and defects between them annihilate in pairs [6, 13]. If it weren't for the product $10 \bullet 01 = 11$, the set $\{00, 01, 10\}$ would form a subalgebra isomorphic to $\star$ (with $00 = 0$, $01 = 1$, and $10 = 1'$) and 11 would not appear after the first time-step.

However, if we choose initial conditions with a single defect, and choose the "phase" of the blocking rule so that we have $\{00, 01\}$ on the left and $\{00, 10\}$ on the right, then the product $10 \bullet 01$ never occurs and the isomorphism is exact. We can use this approach to track multiple defects in parallel, as long as they don't collide. Thus we have

**Corollary.** *The motion of a single defect in rule 18, or of multiple defects that do not collide, can be predicted in* $\mathbf{ACC}^1$.

By tracking single defects until they collide, we should be able to turn this into an algorithm for predicting rule 18 in polylogarithmic parallel time on average. However, there are rare worst cases, where a string of annihilations occur sequentially (each in the light-cone below the previous one) which appear difficult to parallelize. We leave this as an open problem.

## 8  Conclusion

Non-linear cellular automata that are easily predictable are akin to exactly solvable models in statistical mechanics, or integrable non-linear partial differential equations. They help us categorize a hierarchy of dynamical systems between the purely linear and the computationally universal. One might hope that many systems in nature, such as turbulence, reaction-diffusion equations or flocking behavior might, at least in the abstract, be placed somewhere in this hierarchy.

CAs based on groups, semigroups, quasigroups and so on are especially interesting, since they allow us to apply a rich and powerful theory of algebras to the problem of predicting the final state. We have shown here that CAs based on solvable and nilpotent groups can be predicted with $\mathbf{ACC}^1$ and $\mathbf{ACC}^0$ circuits respectively, or in $\mathcal{O}(\log^2 t)$ and $\mathcal{O}(\log t)$ time by an idealized parallel computer whose gates have bounded fan-in. More generally, we have shown that the algebraic CIRCUIT VALUE problem is in $\mathbf{ACC}^1$ for solvable groups, and for a wider class of algebras formed by iterated quasidirect products of commutative semigroups.

We have applied these results to a rule with diffusing defects, which can be used to predict the motion of defects in elementary CA rule 18. It would be very nice if some of the other complex nearest-neighbor CA rules, such as rule 22, 30, or 110, could be analyzed this way; rule 30, for instance, has been proposed as a good cryptographic source of random numbers [31], and a fast prediction algorithm for it would make this use inadvisable. Rule 55 has a more complex chemistry of travelling particles than rule 18 does [7], and it is unclear how to predict their motion efficiently. So far we have been unsuccessful at applying our methods to these CAs.

Can these results be extended to CAs with looser algebraic structure? We note that of the 24 non-isomorphic quasigroups of order 4, for example, 14 are polyabelian and have CAs in $\mathbf{ACC}^0$. The other 10 are capable of expressing arbitrary Boolean functions and so their CIRCUIT VALUE problem (and, we conjecture, their CA prediction problem) is $\mathbf{P}$-complete [21].

We hope these results add to existing $\mathbf{P}$-completeness results for CAs such as majority-voting rules in three or more dimensions [18] as well as other discrete processes such as Ising dynamics and diffusion-limited aggregation [15], and help flesh out this hierarchy from linearity to computational universality.

**Acknowledgements.** C.M. is grateful to Daniel Ashlock, Martin Beaudry, Joshua Berman, Arthur Drisko, Chris Hillman, Derek Holt, Jim Hoover, Peter Johnson, Werner Nickel, Mats Nordahl, Tom Richardson, John Rickard, David Rusin, V. Vinay, and Ross Willard for helpful communications, and to Elizabeth Hunke and Spootie the Cat for companionship. T.P. is grateful to the Santa Fe Institute for an enjoyable visit, although the weather was perhaps unseasonably warm, and there was not a clean glass in the bathroom.

# References

[1] M. Ajtai, "$\Sigma_1^1$ formulae on finite structures." *Ann. Pure Appl. Logic* **24** (1983) 1–48.

[2] David A. Barrington, "Bounded-width polynomial-size branching programs recognize exactly those languages in $\mathbf{NC}^1$." *J. Comput. System Sci.* **38** (1989) 150–164.

[3] M. Beaudry, P. McKenzie, P. Péladeau, and D. Thérien, "Circuits with monoidal gates." Springer LNCS, *Proc. STACS* (1993) 555–565.

[4] W. Burnside, *Theory of Groups of Finite Order.* Dover, 1911.

[5] O. Chein, H.O. Pflugfelder, and J.D.H. Smith, *Quasigroups and Loops: Theory and Applications.* Heldermann Verlag, 1990.

[6] J.P. Crutchfield and J.E. Hanson, "Attractor vicinity decay for a cellular automaton." *Chaos* **3:2** (1993) 215–224.

[7] J.E. Hanson and J.P. Crutchfield, "Computational mechanics of cellular automata: an example." Santa Fe Institute Working Paper 95-10-095, to appear in *Physica* **D**, *Proceedings of the International Workshop on Lattice Dynamics.*

[8] S. Eilenberg and M.P. Schützenberger, "On pseudovarieties." *Advances in Math.* **19** (1976) 413–418.

[9] K. Eloranta, "Partially permutive cellular automata: analysis via tilings subalphabets." Helsinki University of Technology Research report A314 (1992).

[10] M. Furst, J.B. Saxe, and M. Sipser, "Parity, circuits, and the polynomial-time hierarchy." *Math. Syst. Theory* **17** (1984) 13–27.

[11] R. Greenlaw, H.J. Hoover, and W.L. Ruzzo, *Limits to Parallel Computation: P-Completeness Theory.* Oxford University Press, 1995.

[12] M. Hall, *The Theory of Groups.* Chelsea, 1976.

[13] E. Jen, "Exact solvability and quasiperiodicity of one-dimensional cellular automata." *Nonlinearity* **4** (1990) 251.

[14] K. Lindgren and M.G. Nordahl, "Universal computation in simple one-dimensional cellular automata." *Complex Systems* **4** (1990) 299–318.

[15] J. Machta and R. Greenlaw, "The computational complexity of generating random fractals." *J. Stat. Phys.* **82** (1996) 1299

[16] O. Martin, A.M. Odlyzko, and S. Wolfram, "Algebraic properties of cellular automata." *Communications in Mathematical Physics* **93** (1984) 219–258.

[17] C. Moore, "Quasi-linear cellular automata." Santa Fe Institute Working Paper 95-09-078, to appear in *Physica* **D**, *Proceedings of the International Workshop on Lattice Dynamics.*

[18] C. Moore, "Majority-vote cellular automata, Ising dynamics, and P-completeness." Santa Fe Institute Working Paper 96-08-060, submitted to *J. Stat. Phys.*

[19] C. Moore and A. Drisko, "Algebraic properties of the block transformation on cellular automata." Santa Fe Institute Working Paper 95-09-080, submitted to *Complex Systems.*

[20] C. Moore, "Non-Abelian cellular automata." Santa Fe Institute Working Paper 95-09-081.

[21] J. Berman, A. Drisko, F. Lemieux, C. Moore and D. Thérien, "Circuits and expressions with non-associative gates." In progress.

[22] F. Lemieux, *Finite Groupoids and their Applications to Computational Complexity.* Ph. D. Thesis, School of Computer Science, McGill University, Montr'eal (1996).

[23] C.H. Papadimitriou, *Computational Complexity.* Addison-Wesley, 1994.

[24] J. Pedersen, "Cellular automata as algebraic systems." *Complex Systems* **6** (1992) 237–250.

[25] A.A. Razborov, "Lower bounds for the size of circuits of bounded depth with basis $\{\&, \oplus\}$." *Math. Notes Acad. Sci. USSR* **41(4)** (1987) 333–338.

[26] A.D. Robinson, "Fast computation of additive cellular automata." *Complex Systems* **1** (1987) 211–216.

[27] D. Rusin, personal communication.

[28] R. Smolensky, "Algebraic methods in the theory of lower bounds for Boolean circuit complexity." *Proc. 19th ACM Symposium on the Theory of Computing* (1987) 77–82.

[29] M. Suzuki, *Group Theory I.* Springer-Verlag, 1982.

[30] S. Wolfram, "Statistical Mechanics of Cellular Automata." *Reviews of Modern Physics* **55** (1983) 601–644.

[31] S. Wolfram, "Cryptography with Cellular Automata." *Lecture Notes in Computer Science* **218** (1986) 429–432.